\documentclass[acmsmall]{acmart}
\usepackage{pifont}
\usepackage{tikz}
\usepackage{mdframed}
\usepackage{multirow}
\usepackage[normalem]{ulem}
\usepackage{xcolor}

\usepackage{booktabs}      
\usepackage{array}         
\usepackage{threeparttable} 

\usepackage{longtable}     
\usepackage{ltxtable}      
\usepackage{multirow}      
\usepackage{makecell}      

\usepackage[commandnameprefix=always]{changes}  
\definechangesauthor[name={yulei}, color=red]{author1}


\AtBeginDocument{%
  }

\setcopyright{cc}
\setcctype{by}
\acmDOI{10.1145/3808102}
\acmYear{2026}
\acmJournal{PACMSE}
\acmVolume{3}
\acmNumber{FSE}
\acmArticle{FSE095}
\acmMonth{7}
\acmSubmissionID{fse26mainb-p86-p}
\received{2025-09-04}
\received[accepted]{2026-03-24}

\begin{document}

\title[Bash-Commenter: Leveraging Syntax-Aware Preference Optimization to Reinforce Large Language Model...]{Bash-Commenter: Leveraging Syntax-Aware Preference Optimization to Reinforce Large Language Model for Bash Code Comment Generation}

\author{Lei Yu}
\authornote{Affiliated with University of Chinese Academy of Sciences, Beijing, China.}
\orcid{0000-0003-3134-3746}
\affiliation{
  \institution{Institute of Software, Chinese Academy of Sciences}
  \city{Beijing}
  \country{China}
}
\email{yulei2022@iscas.ac.cn}

\author{Jingyuan Zhang}
\authornotemark[1]
\orcid{0000-0001-5475-3815}
\affiliation{
  \institution{Institute of Software, Chinese Academy of Sciences}
  \city{Beijing}
  \country{China}
}
\email{zhangjingyuan2023@iscas.ac.cn}

\author{Xin Wang}
\authornotemark[1]
\orcid{0009-0005-5391-8821}
\affiliation{
  \institution{Institute of Software, Chinese Academy of Sciences}
  \city{Beijing}
  \country{China}
}
\email{wangxin@iscas.ac.cn}

\author{Li Yang}
\authornote{Li Yang and Fengjun Zhang are the corresponding authors.}
\orcid{0000-0001-8364-6525}
\affiliation{
  \institution{Institute of Software, Chinese Academy of Sciences}
  \city{Beijing}
  \country{China}
}
\email{yangli2017@iscas.ac.cn}

\author{Fengjun Zhang}
\authornotemark[2]
\orcid{0000-0002-3830-8786}
\affiliation{
  \institution{Institute of Software, Chinese Academy of Sciences}
  \city{Beijing}
  \country{China}
}
\email{fengjun@iscas.ac.cn}

\author{Peng Wang}
\authornotemark[1]
\orcid{0009-0001-5232-7027}
\affiliation{
  \institution{Institute of Software, Chinese Academy of Sciences}
  \city{Beijing}
  \country{China}
}
\email{wangpeng232@mails.ucas.ac.cn}

\author{Jia Xu}
\authornotemark[1]
\orcid{0009-0004-0143-1707}
\affiliation{
  \institution{Institute of Software, Chinese Academy of Sciences}
  \city{Beijing}
  \country{China}
}
\email{xujia23@mails.ucas.ac.cn}

\author{Jiajia Ma}
\orcid{0000-0002-6028-4186}
\affiliation{
  \institution{Institute of Software, Chinese Academy of Sciences}
  \city{Beijing}
  \country{China}
}
\email{majiajia@iscas.ac.cn}

\renewcommand{\shortauthors}{L. Yu, J. Zhang, X. Wang, L. Yang, F. Zhang, P. Wang, J. Xu, and J. Ma}

\begin{abstract}
Bash script comprehension is a significant challenge in Linux environments due to Bash's syntactic freedom and complex command structures. Despite its critical role in system administration and development, Bash scripts often lack adequate comments, hindering code readability and maintainability. Existing approaches to automated Bash comment generation face two main challenges: (1) Limited training datasets that inadequately represent real-world Bash usage patterns, particularly for complex multi-line scripts; and (2) Insufficient understanding of Bash-specific concepts by Large Language Models (LLMs). Our empirical analysis shows that even after standard training, LLMs still struggle to precisely understand complex Bash command semantics, leading to inaccurate comments. To address these challenges, we propose Bash-Commenter, an advanced comment generation method based on LLaMA-3.1-8B. First, to overcome data limitations (Challenge 1), we construct a comprehensive dataset of complex, multi-line Bash scripts with high-quality comments. Second, to enhance semantic understanding (Challenge 2), we conduct Continual Pre-training (CPT) on large-scale Bash script data, followed by Supervised Fine-tuning (SFT) on our annotated dataset, strengthening the model's foundational knowledge of Bash syntax and semantics. Finally, to resolve the subtle semantic errors that persist, we introduce Syntax-Aware Preference Optimization (SAPO). This method automatically constructs preference pairs by applying single, atomic operations (e.g., modifying a command option or removing an argument) to a script's Abstract Syntax Tree (AST), creating minimal pairs of correct and subtly incorrect scripts. This final optimization stage enables fine-grained command semantics learning and context-dependent quality assessment, significantly improving comment accuracy. We evaluate Bash-Commenter on single-line Bash commands and multi-line Bash scripts. Our method outperforms state-of-the-art baselines, achieving 33.40\% BLEU-4, 58.26\% METEOR, and 57.03\% ROUGE-L for 1,064 single-line commands, and 22.15\% BLEU-4, 43.89\% METEOR, and 32.80\% ROUGE-L for 1,046 multi-line scripts. Moreover, human evaluation and LLM evaluation demonstrate the superior quality of comments generated by Bash-Commenter in terms of correctness, completeness, and naturalness.
\end{abstract}


\maketitle

\section{Introduction}

Bash, the default domain-specific language (DSL) in Linux systems for tasks like file management, network control, and process management \cite{newham2005learning}, remains indispensable for Linux system development and maintenance despite its narrower application range compared to general-purpose programming languages (GPPLs) \cite{lin2018nl2bash}. Known for its high flexibility and syntactical freedom, Bash is a powerful tool, but these very characteristics also pose significant challenges for code comprehension \cite{lin2018nl2bash}. This challenge is particularly pronounced for developers unfamiliar with Bash, who often struggle to understand the functionality and semantics of Bash scripts. Based on our statistical analysis of Stack Overflow, as of March 10, 2025, there are 93,046 Q\&A posts related to the keyword “shell” and 156,721 Q\&A posts related to the keyword "bash" highlighting the urgent need for tools to assist in understanding Bash scripts \cite{StackOverflow_bash, StackOverflow_shell}. Studies suggest that developers spend up to 59\% of their time on code comprehension, and high-quality comments are critical for improving code readability and maintainability \cite{xia2017measuring, lu2023llama}. Many developers often fail to provide sufficient comments due to oversight or time constraints, which leads to missing or outdated annotations \cite{shen2024bash, lin2018nl2bash}. Longitudinal analysis of Stack Overflow activity (1134 days post-ChatGPT release~\cite{stackoverflow_bash_1134days} vs 3000 day baseline~\cite{stackoverflow_bash_3000days}, collected January 7, 2026) shows Bash retains 14.39\% of baseline volume, 6.54\% relatively higher than SQL and 18.07\% relatively higher than Java, indicating LLMs struggle more with Bash-specific syntax than general programming. This persistent demand stems from Bash's safety-critical nature: unlike GPPLs, Bash's syntactic freedom allows direct OS manipulation where semantic misinterpretation (e.g., quoting in \texttt{rm -rf "\$VAR"} vs \texttt{rm -rf \$VAR}, block sizes in \texttt{dd} commands) can cause irreversible data loss. High-quality Bash comments serve four critical applications: (1) DevOps maintenance (where developers spend 59\% of time on comprehension~\cite{xia2017measuring}), (2) security auditing to identify risky operations (recursive deletion, permission changes, network operations), (3) developer assistance for shell-related posts, and (4) legacy code modernization: comprehending undocumented infrastructure scripts for safe migration to containerized platforms or rewriting in modern languages. This challenge reflects a broader issue for domain-specific languages (DSLs). Unlike general-purpose programming languages (GPPLs) like Java and Python where automated comment generation is well-studied\cite{lu2023llama, geng2024large}, DSLs such as Bash remain largely unexplored in this context \cite{yu2024smart, yu2023pscvfinder, yu2022bashexplainer}, despite an urgent need to address this gap.

Recent work on Bash code comment generation has explored advanced techniques to enhance the quality and relevance of generated comments. Yu et al. \cite{yu2022bashexplainer} proposed the BASHEXPLAINER framework, which employs a two-stage training strategy. In the first stage, CodeBERT is used to semantically encode Bash code, while the second stage integrates an information retrieval module with a deep learning-based generator to produce comments by leveraging semantic and lexical similarity. Shen et al. \cite{shen2024bash} introduced Bash2Com, which further enhanced comment generation through adversarial training (NP-GD) and a semantic-aware module (MASA). More recently, Zhang et al. \cite{zhang2025bash} proposed HBCom, which leverages a Heterogeneous Information Graph (HIG) to integrate syntactic and semantic features, providing a more comprehensive understanding of Bash commands.

\begin{figure*}[htbp]
\centerline{\includegraphics[width=0.9\textwidth]{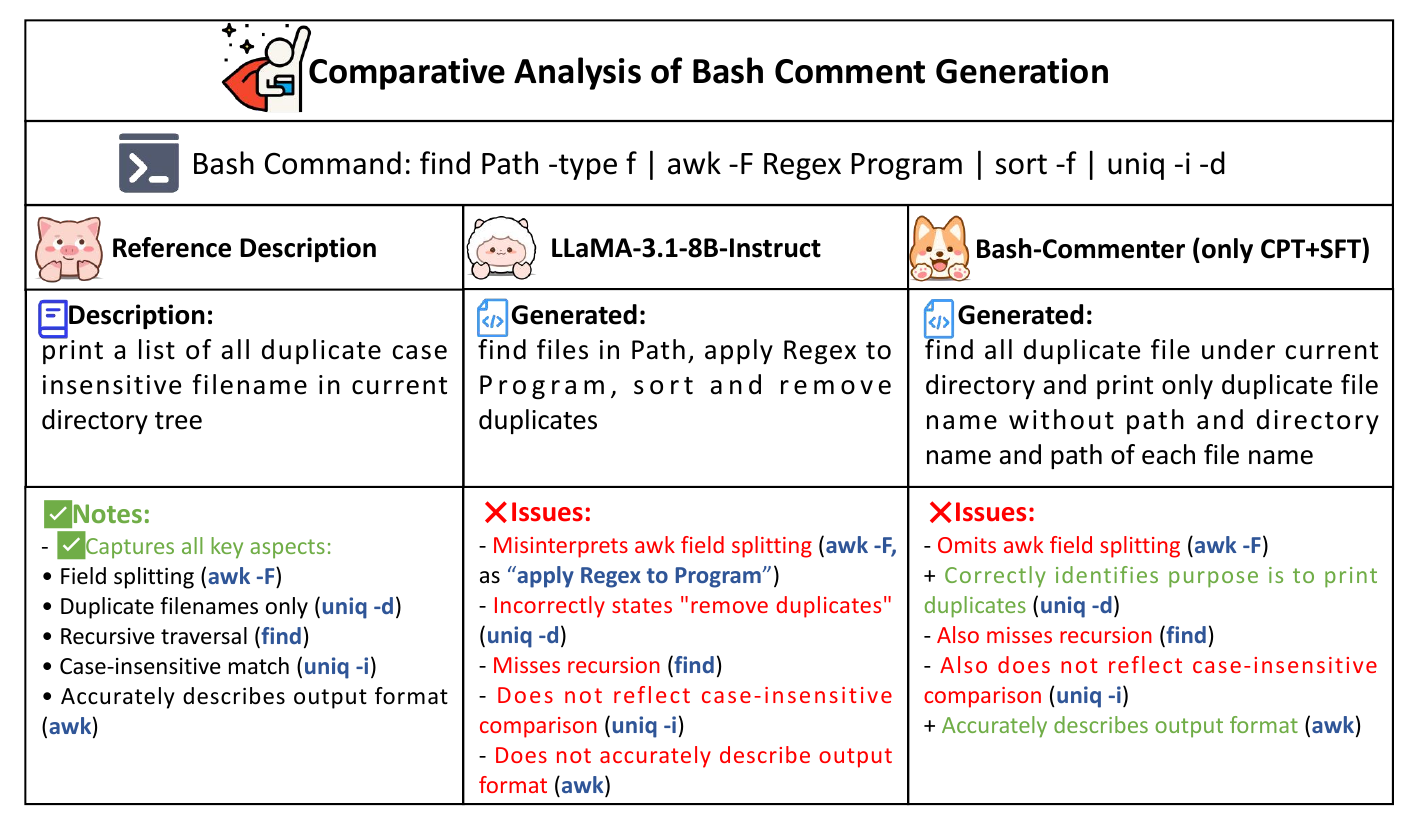}}
\caption{A motivation example to illustrate the limitations of LLM-based Bash comment generation, motivating the need for SAPO.}
\label{tab:motivation}
\end{figure*}

Recently, Large Language Models (LLMs) have attracted significant interest due to their exceptional ability to understand and generate natural language. They have shown outstanding efficacy across a range of complex applications, such as code comprehension and generation \cite{liu2024your, weyssow2023exploring, lu2023llama, cheng2025auvana, shen2024dependency, zan2024swe, yu2026sql}, as well as smart contract auditing and generation \cite{yu2024smart, yu2025smart, yu2023pscvfinder, yu2025sael, yuan2025mos, yu2025towards, yuan2025leveraging}. General LLMs show promise in adapting to new patterns \cite{naveed2023comprehensive}. However, they often encounter significant difficulties in handling Bash scripting-specific concepts and command execution nuances. As shown in Fig. \ref{tab:motivation}, when comparing comments from a reference description, a general LLM (LLaMA-3.1-8B-Instruct), and a specialized model after continual pre-training and supervised fine-tuning (Bash-Commenter with CPT+SFT), both models still exhibit critical misunderstandings. The general LLM fundamentally misinterprets multiple aspects of the command, including the \texttt{awk} field splitting mechanism, the purpose of \texttt{uniq -i}, and the recursive nature of \texttt{find}. Bash-Commenter (CPT+SFT), while improving in certain aspects such as correctly identifying the main purpose of finding duplicate files (\texttt{uniq -d}) and accurately describing the output format (\texttt{awk}), still presents significant deficiencies. It continues to omit the recursive traversal functionality of the \texttt{find} command, fails to accurately express the role of case-insensitive matching (\texttt{uniq -i}), and omits the field splitting functionality of \texttt{awk}. These persistent errors demonstrate that traditional training approaches (CPT+SFT) are insufficient for developing models with accurate understanding of complex Bash commands.

To address these challenges, we propose \textbf{Bash-Commenter}, a 
specialized model based on the LLaMA-3.1-8B model, developed through a 
comprehensive three-stage pipeline. \textbf{First}, to address the data scarcity and quality issues (as detailed in Section~\ref{sec:motivation}), we curate a new high-quality dataset sourced from the large-scale empirical study by Dong et al.~\cite{dong2023bash}. Unlike previous datasets that mainly consist of single-line commands, our dataset features more complex and multi-line scripts with detailed annotations, better reflecting the practical needs of Bash code comment generation in real-world scenarios. \textbf{Second}, to overcome the limitations of general LLMs in understanding Bash-specific syntax and semantics (as empirically demonstrated in Section~\ref{sec:motivation}), we utilize large-scale Bash script data for domain-specific Continual Pre-training (CPT), followed by Supervised Fine-tuning (SFT) on our curated dataset. This two-step process enhances the model's foundational knowledge of Bash commands and its capability to annotate complex multi-line scripts. \textbf{Finally}, to resolve the subtle but critical issues in comment quality that persist even after CPT and SFT (also noted in Section~\ref{sec:motivation}), we introduce our primary 
innovation: \textbf{Syntax-Aware Preference Optimization (SAPO)}. Instead of using generic preference pairs for Direct Preference Optimization (DPO), SAPO employs a fully automated pipeline to generate "Minimal Syntactic Pairs" by applying single "atomic operations" to a command's Abstract Syntax Tree (AST). This targeted approach transforms DPO into a "knowledge injection" process, achieving the goal of \textbf{Fine-grained Command Semantics Learning} by forcing the model to master the precise semantics of command-specific details, such as correctly recognizing \texttt{awk} field splitting, \texttt{find}'s recursive traversal functionality, and \texttt{uniq -i}'s case-insensitive comparisons. This process also enables \textbf{Context-dependent Quality Assessment}, learning to provide the most relevant comment based on specific analysis needs as demonstrated in Fig. ~\ref{fig:bash-command3}, ensuring the final comments are not only technically accurate but also conceptually insightful.

In the data construction process, we handled single-line and multi-line Bash scripts separately. For single-line commands, SFT data was drawn from the BASHEXPLAINER dataset \cite{yu2022bashexplainer}, with typographical errors corrected. For multi-line scripts, initial comments were generated by DeepSeek-V3-0324 and Qwen2.5-Max, scored by DeepSeek-R1 for correctness, completeness, and naturalness, and then reviewed and refined by bash experts (average over 5 years Linux experience). For the SAPO dataset, we designed a fully automated pipeline to generate "Minimal Syntactic Pairs." This method systematically applies single, atomic operations (e.g., modifying a command option or removing an argument) to a script’s Abstract Syntax Tree (AST), creating a pair consisting of the original script and a subtly incorrect variant. Specifically, we programmatically modify a single node within the AST and then reconstruct the script string from the modified tree; scripts that fail the initial parsing are discarded to ensure robustness. To form the preference pair, the 'chosen' comment and the 'rejected' comment were both generated by a single powerful LLM, DeepSeek-V3-0324. Unlike the multi-model voting process in the SFT stage, this single-model approach was intentionally used to ensure stylistic consistency, thereby forcing the model to learn the precise semantic difference caused by the atomic operation rather than superficial stylistic variations.

Experimental results on 1,064 single-line Bash commands and 1,046 multi-line Bash scripts confirm Bash-Commenter's superior performance. It achieved the best performance (BLEU-4, METEOR, ROUGE-L): 33.40\%, 58.26\%, 57.03\% for single-line, and 22.15\%, 43.89\%, 32.80\% for multi-line scripts. LLM evaluation show positive ratings (score 3 or 4) for correctness, completeness, and naturalness: for single-line commands (76.5\%, 80.0\%, 86.5\%) and multi-line scripts (70.0\%, 68.0\%, 73.0\%). Human evaluation further underscore its quality, with positive ratings (score 3 or 4) for correctness, completeness, and naturalness: for single-line commands (79.5\%, 82.0\%, 89.0\%) and multi-line scripts (73.0\%, 71.0\%, 76.0\%), significantly outperforming baselines.

The main contributions of this paper are as follows:
\begin{itemize}
\item To the best of our knowledge, we are the first to introduce Syntax-Aware Preference Optimization in the bash code comment generation task.
\item We constructed and publicly released comprehensive datasets for training and evaluating Bash comment generation, uniquely featuring complex multi-line scripts.
\item We propose Bash-Commenter, a new method combining CPT, SFT and SAPO for bash code comment generation, achieving state-of-the-art performance through automatic metrics, LLM evaluation, and human evaluation.


\end{itemize}

\section{Background and Motivation}


\subsection{Problem Statement}

We propose an automated approach for Bash script comment generation. Given a Bash command or script $x$, our system generates a descriptive comment $\hat{y}$ that accurately captures the script's functionality. The task is formalized as finding an optimal mapping function $f: X \rightarrow Y$ where $X$ represents the space of Bash scripts and $Y$ represents comments.

\subsection{Motivations}
\label{sec:motivation}

In this section, we analyze the motivations behind our research on bash code comment generation. 

\begin{table*}[ht]
\setlength{\tabcolsep}{1.0pt}
\centering
\caption{Comparison of BASHEXPLAINER and Bash-Commenter Training Datasets}
\label{tab:dataset-comparison}

\begin{tabular}{lccccccccc}
\toprule
Dataset & Samples & SL/ML (\%) & AvgLen & AvgLn & CmtLen & Cmds & P/R (\%) & DistCmds & Anno \\
\midrule
BASHEXP \cite{yu2022bashexplainer}  & 8,469   & 100.0/0.0    & 8.5   & 1    & 12    & 1.5  & 40.5     & 466      & Expert \\
Ours  & 18,657  & 49.4/50.6   & 54.6  & 17.2 & 123   & 9.7  & 55.8     & 17,122   & LLM+Exp \\
\bottomrule
\end{tabular}
\noindent
\footnotesize
\textbf{Notes:} SL/ML: Single-line Commands / Multi-line Scripts (\%); SL/ML shows the percentage ratio of single-line commands to multi-line scripts; AvgLen/AvgLn: avg tokens/lines; CmtLen: avg comment length; Cmds: commands per sample; P/R: pipe/redirection (\%); DistCmds: distinct commands; Anno: annotation type.
\end{table*}


Following the taxonomy proposed by Chen et al.~\cite{chen2021my} 
and Geng et al.~\cite{geng2024large}, code comments serve two distinct 
purposes: (1) \textbf{Descriptive comments} explain command-level 
functionality and parameters (e.g., ``-mtime +7 finds files modified 
more than 7 days ago''), and (2) \textbf{Intent-level comments} 
articulate high-level goals (e.g., ``This script automates log rotation 
to manage disk space''). For Bash, this naturally maps to script 
complexity: \textbf{single-line commands (49.4\%) primarily need 
descriptive comments} for syntax and parameters, while \textbf{multi-line 
scripts (50.6\%) need both types}, with descriptive comments for 
individual commands and intent-level comments for overall workflow. 
As shown in Table~\ref{tab:dataset-comparison}, this dual-category 
design enables Bash-Commenter to generate both precise parameter 
explanations and comprehensive workflow documentation.

\textbf{Motivation 1: Limitations of Existing Bash Code-Comment Dataset.}
Previous research on Bash code comment generation has focused almost exclusively on single-line Bash commands, resulting in datasets that are short, simple, and fail to capture the complexity of real-world scripting. For example, as shown in TABLE \ref{tab:dataset-comparison}, the BASHEXPLAINER dataset \cite{yu2022bashexplainer} consists entirely of single-line commands (100\%), with an average length of only 8.5 tokens, average comment length of 12 tokens, and just 1 line per sample. It covers only 466 distinct Bash commands and 40.5\% of samples contain pipe or redirection, making it difficult for models to generalize to practical, professional scenarios. The BASHEXPLAINER dataset is also used by recent works such as Bash2Com~\cite{shen2024bash} and HBCom~\cite{zhang2025bash}.

To better reflect real-world scripting practices, our dataset combines two sources with distinct annotation strategies: \textbf{(1) Single-line commands (reused with corrections):} All 8,469 commands from BASHEXPLAINER \cite{yu2022bashexplainer}, with factual errors (e.g., "directori"→"directory") and syntactic mistakes corrected before SFT (Section \ref{sec:sft_data_construction}). \textbf{(2) Multi-line scripts (newly annotated):} 8,154 scripts sampled from Dong et al. \cite{dong2023bash}, for which we generated and expert-verified all new annotations via LLM+expert pipeline (Section \ref{sec:sft_data_construction}). The construction process was built upon the large-scale empirical study of Dong et al. \cite{dong2023bash}, which demonstrated that multi-line scripts and complex structures are common in professional Bash development. Inspired by this finding, we excluded auto-generated and trivial scripts from Dong et al. \cite{dong2023bash} (e.g., scripts with fewer than 3 lines, fewer than 20 tokens, or containing patterns like "DO NOT EDIT"), and further removed near-duplicates using a Jaccard Index threshold of 0.9. We then sampled a diverse range of examples from domains such as system administration, DevOps, data processing, and network management to ensure practical coverage. \textbf{(3) SAPO preference data:} All 2,034 preference pairs are constructed from the same code sources as SFT (748 single-line, 1,286 multi-line). Chosen comments for single-line data reuse corrected BASHEXPLAINER annotations; all other annotations (chosen comments for multi-line and all rejected comments) are newly generated via AST-based mutation (Section \ref{sec:sapo_data}).


As shown in TABLE \ref{tab:dataset-comparison}, the resulting Bash-Commenter 
dataset is much larger and more representative of real-world Bash usage than 
previous datasets. It contains 18,657 samples, of which half (50.6\%) are 
multi-line scripts. The average sample length reaches 54.6 tokens, with 17.2 
lines and 9.7 Bash commands per sample. Notably, 55.8\% of samples involve pipes 
or redirection, and the dataset includes 17,122 distinct Bash commands. The 
average comment length is as high as 123 tokens, and for single-line data, annotations are expert-written with our corrections; for multi-line data, all annotations are LLM-generated and expert-verified (Section \ref{sec:sft_data_construction}).

\begin{figure*}[t]
\centering
\begin{minipage}{\linewidth}
\centering
{\small\ttfamily
find /data -type f -name "*.log" | grep -i "error" | awk '\{print \$1,\$5\}' > errors\_summary.txt
}
\end{minipage}
\caption{Example of a Bash command pipeline for error log processing.}
\label{fig:bash-command3}
\end{figure*}

\textbf{Motivation 2: Limitations in Comment Quality of LLM-based Bash Comment Generation Methods.} As shown in Fig. \ref{tab:motivation}, general LLMs (LLaMA-3.1-8B-Instruct) fail in two aspects: they fundamentally misinterpret multiple aspects of Bash commands and provide inaccurate comment. Although specialized LLMs like Bash-Commenter (trained through CPT and SFT) perform better in command analysis, they still struggle to provide fully accurate comment. The specialized LLM correctly identifies the main purpose of finding duplicate files (\texttt{uniq -d}) and accurately describes the output format (\texttt{awk}). However, its comment contains critical misunderstandings of command functionality. It omits the recursive traversal functionality of the \texttt{find} command, fails to accurately express the role of case-insensitive matching (\texttt{uniq -i}), and overlooks the field splitting functionality of \texttt{awk}. This limitation in comment quality persists despite the improvements from CPT and SFT. It underscores the need for further refinement in the LLM's understanding of Bash command execution flow and operational semantics. While general LLMs like LLaMA-3.1-8B-Instruct struggle with Bash semantics (Fig.~\ref{tab:motivation}), this limitation extends to commercial models. GPT-4.1 and Claude-3.7-Sonnet achieve only 11.89\% and 11.50\% BLEU-4 on multi-line scripts respectively (Table~\ref{multiline_comparison}), with human evaluations showing 28\% and 27\% perfect correctness ratings (Table~\ref{evaluation of explanation}). Case analysis (Fig.~\ref{case_study}) reveals systematic failures across all methods in recognizing safety mechanisms (\texttt{-print0}/\texttt{xargs -0}) and regex types (\texttt{posix-egrep}).


Unlike traditional training methods that rely on absolute labels, SAPO learns from relative preferences between paired comments which is especially effective in two key aspects:

(1) \textbf{Fine-grained Command Semantics Learning:} SAPO's loss function explicitly models the relative preference between two comment, enabling the capture of subtle command-specific details in Bash scripts. As illustrated in Fig. \ref{tab:motivation}, even Bash-Commenter (CPT+SFT) struggles with critical issues such as omitting \texttt{awk} field splitting details, misunderstanding recursion with \texttt{find}, and failing to recognize case-insensitive comparisons. Our SAPO approach addresses these limitations by learning from human preferences between command-level comments. By training on paired examples where experts prefer correct interpretations over incorrect ones, the model gradually learns to distinguish subtle aspects of command behavior—properly recognizing when \texttt{awk -F} indicates field splitting rather than regex application, understanding that \texttt{find} implies recursive traversal by default, and acknowledging the case-insensitive nature of \texttt{uniq -i} operations. This preference-guided learning creates a more nuanced understanding of command semantics, critical for avoiding potentially catastrophic errors in practical scripting scenarios.

(2) \textbf{Context-dependent Quality Assessment:} Bash commands often involve complex interactions between utilities and data transformations, making comment quality highly context-dependent. SAPO's preference learning effectively captures how comment quality varies with context through paired comparisons. For example, consider the Bash command in Fig. \ref{fig:bash-command3} with two different comments:

\textbf{Comment A:} "This command recursively searches for all \texttt{.log} files in the \texttt{/data} directory, filters lines containing \texttt{'error'} (case-insensitive), extracts the \texttt{1st} and \texttt{5th} columns from each matching line, and saves the results to \texttt{errors\_summary.txt}"


\textbf{Comment B:} "This command analyzes logs by finding all log files in the data directory tree and filtering for error messages using case-insensitive matching. It extracts timestamp and error message columns, then saves these details to a summary file."

While both describe the same command, their value differs by context. Comment A suits technical references with precise syntax descriptions, while Comment B fits higher-level documentation like user guides by focusing on conceptual purpose and usage patterns. Through such paired comparisons, SAPO learns context-appropriate comment generation, addressing limitations of CPT and SFT-only training.

\section{Approach}

\begin{figure*}[htbp]
\centerline{\includegraphics[width=0.9\textwidth,height=0.4\textheight]{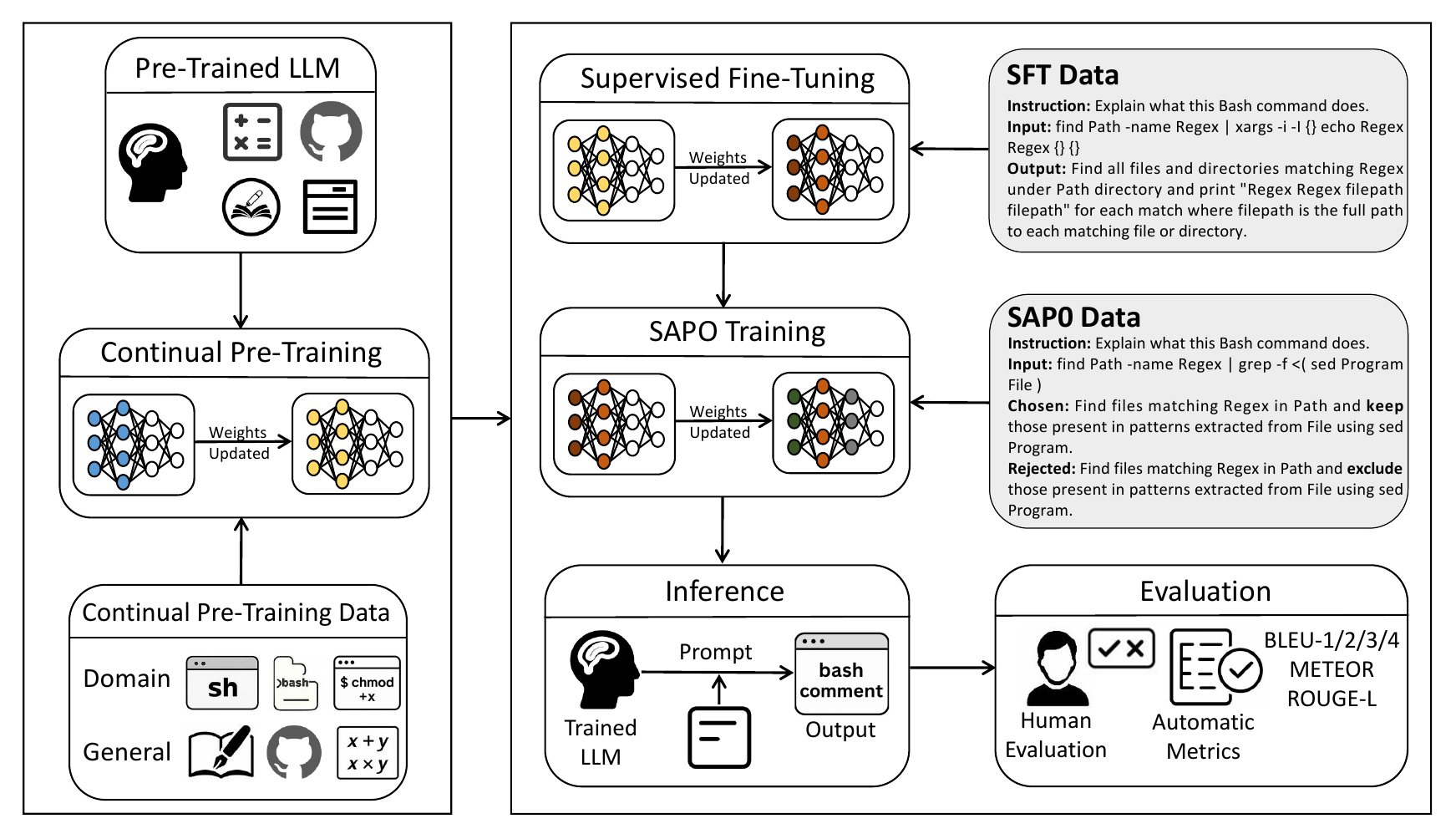}}
\caption{The Overview of our Bash-Commenter.}
\label{tab: overview}
\end{figure*}
\begin{figure}[!h]
\centerline{\includegraphics[width=1\textwidth]{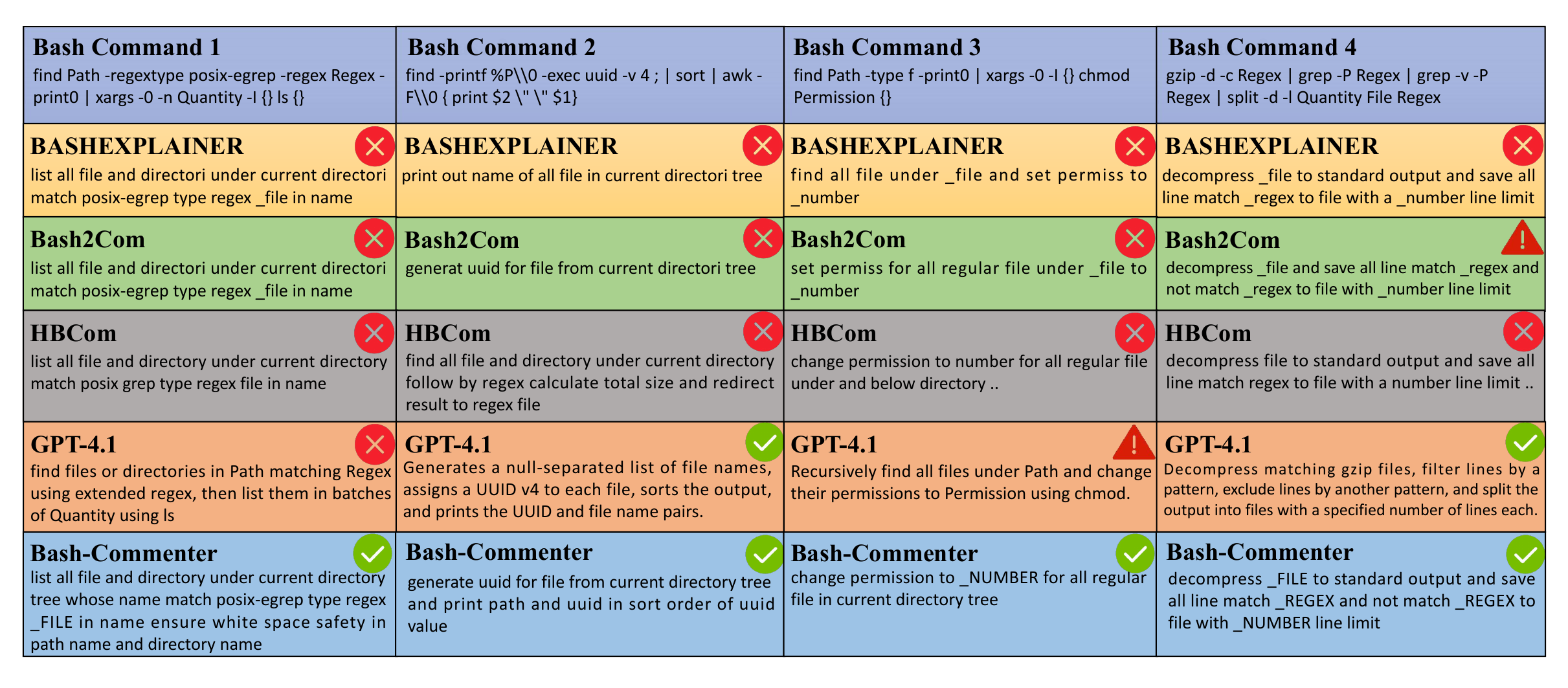}}
\caption{Case Study of Bash Code Comment Generation Using Bash-Commenter.}
\label{case_study}
\end{figure}

Bash-Commenter follows three stages (Fig.\ref{tab: overview}): CPT, SFT, and SAPO, built on LLaMA-3.1-8B for its optimal performance-efficiency balance\cite{lu2023llama, yu2024smart, tang2025breaking}. \textbf{Design Rationale.} SAPO-based fine-tuning is prioritized over alternatives for three reasons: (1) \textit{Retrieval fails on semantics}: RAG baselines like BASHEXPLAINER reach only 29.13\% BLEU-4 (Table~\ref{single_line_comparison}), struggling to distinguish subtle command opposites (e.g., \texttt{grep -v} vs. \texttt{grep}), while BM25/VSM performs worse (9.40\%--19.24\%). (2) \textit{Prompting misses syntax nuances}: even GPT-4.1 overlooks critical option combinations and pipeline interactions (Fig.\ref{case_study}). (3) \textit{Agents are inefficient}: multi-agent systems incur high latency without guaranteeing AST-level correctness. SAPO instead directly embeds syntax rules into model weights via AST-based minimal pairs, enabling efficient single-pass inference. LLaMA-3.1-8B (July 2024) further supports practical deployment, outperforming comparable models and rivaling those 4×4× larger (Table\ref{single_line_comparison}).


\subsection{Open-Source Bash Scripts Collection}

\subsubsection{Continual Pre-training}

For our Continual Pre-training dataset, we followed the approach in \cite{dong2023bash} and integrated Bash-related data from multiple sources: (1) The complete collection of Linux manual pages (Man Pages), containing detailed command descriptions and usage examples; (2) Stack Overflow question-answer pairs tagged with "bash", "shell", and "linux"; (3) Bash projects from the top 1,000 GitHub repositories by star count (following \cite{dong2023bash}); (4) Curated shell script discussions from Unix \& Linux Stack Exchange. We extracted  a total of 1,013,135 Bash commands and scripts from these sources. To ensure uniqueness, scripts were filtered using a Jaccard Index similarity threshold of 0.9, eliminating those with over 90\% token similarity. After deduplication, 676,524 unique Bash scripts remained. The threshold of 0.9 was adopted following prior work \cite{storhaug2023efficient, yu2024smart}, which used this value to identify and remove near-duplicate code snippets.

\subsubsection{Supervised Fine-Tuning and Syntax-Aware Preference Optimization}

\textbf{For SFT dataset}, we combined data from two primary sources: the BASHEXPLAINER dataset \cite{yu2022bashexplainer} and selected scripts from \cite{dong2023bash}. The BASHEXPLAINER dataset \cite{yu2022bashexplainer} provided 8,469 professionally annotated single-line Bash commands with high-quality comment. To augment this dataset, we incorporated an additional 8,154 annotated multi-line Bash scripts from \cite{dong2023bash}, focusing on diverse use cases and complexity levels. After quality control and deduplication, our final SFT dataset contained 16,623 high-quality $\langle$Bash Code, Comment$\rangle$ pairs. Our quality control employed a four-stage pipeline: (1) Jaccard-based deduplication (threshold 0.9); (2) LLM scoring (DeepSeek-R1) with 50/30/20 weighting for correctness/completeness/naturalness, filtering samples $\geq$ 7.0/10 and excluding 23.4\% low-quality candidates; (3) expert verification by 24 professionals (dual annotation on 300 samples, Fleiss' Kappa=0.78); (4) re-annotation for confidence <6, with 8.7\% requiring arbitration (details in Sections \ref{sec:llm_scoring} - \ref{sec:sft_data_construction}). \textbf{For SAPO dataset}, we constructed paired data comprising preferred and rejected comment for the same Bash commands or Bash scripts. We leveraged 748 pairs from \cite{yu2022bashexplainer}. Additionally, we created 1,286 preference pairs from the \cite{dong2023bash} collection. We ensured there was no overlap between the SAPO dataset and the SFT dataset. In total, our SAPO dataset consisted of 2,034 preference pairs from single-line commands to multi-line scripts.


\subsection{SFT and SAPO Data Construction}

\subsubsection{Initial Data Generation}

We utilized DeepSeek-V3-0324 \cite{liu2024deepseek} and Qwen2.5-Max \cite{yang2024qwen2} for initial annotation generation, chosen for their performance comparable to models like GPT-4.1 and Claude-3.7-Sonnet at a lower cost, as indicated by their technical reports \cite{liu2024deepseek,yang2024qwen2}. The Bash scripts were collected from GitHub, Stack Overflow, and Linux system administrator communities \cite{dong2023bash}. For each script, corresponding functional comment annotations were generated.

\subsubsection{LLM Scoring}
\label{sec:llm_scoring}


DeepSeek-R1 served as the judge model for scoring generated comments, selected for its strong reasoning, code evaluation performance, and cost-effectiveness~\cite{guo2025deepseek}. It assigned per-aspect scores (1--10) across Correctness, Completeness, and Naturalness (Section~\ref{subsubsec:evaluation_criteria}), and computed a weighted overall score (Correctness 50\%, Completeness 30\%, Naturalness 20\%) following Smart-LLaMA-DPO~\cite{yu2025smart}, which prioritizes correctness for code explanation. Only annotations exceeding a weighted score of 7.0 advanced to human verification, automatically filtering out 23.4\% of initial candidates. Empirical validation (Table~\ref{evaluation of explanation}) confirms this weighting: when correctness=1, 85.7\% of samples (6/7) exhibit low completeness (scores 1--2), while correctness=3--4 samples consistently achieve high scores across all metrics. The 7.0/10 threshold follows prior work~\cite{yu2025smart}.

\subsubsection{SFT Data Construction}
\label{sec:sft_data_construction}

For multi-line scripts, annotations with higher LLM scores were selected and refined by 24 Bash experts (12 senior Linux system administrators averaging 8 years of experience, 7 DevOps engineers averaging 6 years, and 5 computer science researchers averaging 5 years), who verified command functionality accuracy, improved parameter comments, and ensured clarity and completeness. Inter-annotator consistency over 300 dual-annotated samples (200 single-line, 100 multi-line) yielded Fleiss' Kappa = 0.78. Each annotation received a confidence score (1--10); those below 6 were independently re-annotated by a second expert with consensus discussion, and unresolved cases were arbitrated by a third expert (8.7\% arbitration rate), ensuring all final annotations exceeded confidence 6, with 89.3\% exceeding 7. Independent quality assessment by 5 senior experts (5+ years) on 500 stratified samples (247 single-line [49.4\%], 253 multi-line [50.6\%]) yielded 8.3/10 (Correctness), 8.1/10 (Completeness), and 8.5/10 (Naturalness), with a 91.2\% approval rate (Correctness $>$ 7). Error analysis of the remaining 44 samples (8.8\%) revealed 4.4\% minor wording issues, 3.0\% incomplete parameter explanations, and 1.4\% factual errors (removed). For single-line commands, spelling errors in BASHEXPLAINER~\cite{yu2022bashexplainer} (e.g., directori→directory) were corrected. The final SFT dataset comprises 16,623 high-quality samples. Following human-LLM collaborative annotation practices~\cite{yu2025smart}, this demonstrates that LLM-generated labels with systematic human refinement can match human-written annotation quality. Validation on 500 randomly sampled BASHEXPLAINER commands re-annotated via our methodology, blindly rated by 5 senior experts (5+ years, Fleiss' Kappa = 0.76), showed no significant differences across all metrics (Correctness: 8.6±1.2 vs. 8.5±1.1, p=0.421; Completeness: 8.3±1.4 vs. 8.4±1.2, p=0.587; Naturalness: 8.7±1.1 vs. 8.8±0.9, p=0.312), confirming statistical equivalence.

\subsubsection{Automated Generation of Minimal Syntactic Pairs for SAPO}
\label{sec:sapo_data}

To enable our Syntax-Aware Preference Optimization (SAPO) framework, we designed and implemented a \textbf{fully automated data generation pipeline}. This pipeline addresses a key limitation of standard DPO: the reliance on generic or manually created preference pairs, which provide diffuse and inefficient learning signals. Our approach, instead, systematically generates \textbf{``Minimal Syntactic Pairs''} to provide the model with highly targeted, interpretable, and scalable feedback.

We analyzed 856 Stack Overflow posts (January 2020 to November 2022, pre-LLM era to ensure human-verified corrections) with six coders (Fleiss' $\kappa$ = 0.82), identifying five atomic operations covering 73.4\% of errors: Option Deletion (28.7\%), Argument Modification (22.1\%), Command Replacement (12.6\%), Redirection Removal (6.0\%), Pipeline Adjustment (4.0\%).

\textbf{Formal Definition of a Minimal Syntactic Pair:} A preference tuple \texttt{<C, chosen, rejected>} is defined as a ``Minimal Syntactic Pair'' if it satisfies the following conditions:
\begin{enumerate}
    \item \textbf{Semantic Anchor:} The \textit{chosen} comment is a correct and high-quality description of an original Bash command \texttt{C}.
    \item \textbf{Syntactic Perturbation:} The \textit{rejected} comment is a correct description of a variant command \texttt{C'}, which is derived from \texttt{C} by applying a single, predefined \textbf{``Minimal Atomic Operation''} to its Abstract Syntax Tree (AST).
    \item \textbf{Surface Similarity:} The \textit{chosen} and \textit{rejected} comments are naturally similar in language, isolating the semantic difference caused exclusively by the atomic operation.
\end{enumerate}


A “Minimal Atomic Operation” is a single, targeted transformation on the command’s AST that corresponds to a common but critical developer error. These atomic operations were not chosen arbitrarily. Based on a preliminary manual analysis of real-world Bash errors collected from Stack Overflow and the failure cases of our initial SFT model, we developed a wide range of such atomic operations to target the most critical error patterns. For brevity, we present and analyze five of the most representative categories, which are detailed as examples in Table~\ref{tab:atomic_operations}.


\begin{table}[h]
\centering
\caption{Examples of Minimal Atomic Operations for SAPO Data Generation}
\label{tab:atomic_operations}
\begin{tabularx}{0.95\linewidth}{l >{\raggedright\arraybackslash}X >{\raggedright\arraybackslash}X} 
\toprule
\textbf{Atomic Operation} & \textbf{Description} & \textbf{Example \texttt{C -> C'}} \\
\midrule
Option Deletion & Removes a command-line option (flag). & \texttt{grep -v "a" -> grep "a"} \\
Option Substitution & Replaces an option with a different one. & \texttt{head -n 5 -> tail -n 5} \\
Argument Mod. & Modifies an argument (\textit{e.g.}, quote type). & \texttt{echo "\$VAR" -> echo '\$VAR'} \\
Redirection Del. & Removes a file redirection operator. & \texttt{ls > f.txt 2>\&1 -> ls > f.txt} \\
Pipeline Simp. & Removes a component from a pipeline. & \texttt{sort | uniq -c -> sort | uniq} \\
\bottomrule
\end{tabularx}
\end{table}


\textbf{The Automated Generation Pipeline:} Our pipeline consists of four automated stages:
\begin{enumerate}

    \item \textbf{Command Variant Generation:} For each source command \texttt{C}, we first use a robust Bash parser (\texttt{bash-parser} for Python) to generate its AST. To create the variant command \texttt{C'}, we then apply one of the predefined atomic operations, which involves programmatically modifying a single node within the AST and then reconstructing the script string from the modified tree. Any command that fails this initial parsing step is automatically excluded.
    
    \item \textbf{Parallel Annotation Generation:} We use a powerful LLM (DeepSeek-V3-0324) to generate annotations for both commands in parallel. The annotation for \texttt{C} becomes the candidate \textit{chosen} comment, and the annotation for \texttt{C'} becomes the candidate \textit{rejected} comment.
    
    \item \textbf{SAPO Sample Assembly:} The final SAPO training sample is assembled as \texttt{<prompt=C, chosen=Comment\_C, rejected=Comment\_C'>}. This structure forces the model to learn why \texttt{Comment\_C'} is incorrect \textit{for the original command \texttt{C}}, thereby learning the specific semantics of the atomic operation.

    \item \textbf{LLM-as-a-Judge Verification:} To ensure quality without human intervention, we employ an LLM-as-a-Judge mechanism (using DeepSeek-R1). Each generated triplet is automatically verified for accuracy and integrity, with any that fail this check being discarded to ensure the final dataset's quality. To validate our judge, we manually reviewed a random sample of 200 of its decisions and found a 96\% agreement rate (192 out of 200) with human experts.
\end{enumerate}

\subsection{Continual Pre-Training}


To develop Bash-Commenter, we continual pre-train LLaMA-3.1-8B for two epochs on Bash-related data and one epoch on natural language data, enhancing both code and language capabilities. We optimize the language modeling objective used in GPT and LLaMA. Given a sequence $x$ of $n$ tokens $t_0, t_1, ..., t_{n-1}$, we maximize the likelihood of the entire sequence by calculating the product of conditional probabilities for each token:

\begin{equation}
p(x) = \prod_{i=1}^{n-1} p(t_i|t_1, t_2, ...t_{i-1})
\end{equation}

\textbf{Pre-training Dataset [809.79M tokens]}. To build a comprehensive and efficient pre-training dataset, we sampled high-quality data of two main types: \textbf{(1) Bash-related data} [554.40M tokens]. To enhance Bash comment generation, we gathered Bash scripts and commands from diverse sources \cite{dong2023bash}: Linux manual pages, Stack Overflow Q\&A tagged with "bash", "shell", and "linux", highly-starred Bash projects on GitHub, and curated discussions from Unix \& Linux Stack Exchange. This collection forms a comprehensive Bash code repository, spanning simple commands to complex scripts. \textbf{(2) Other related data} [255.39M tokens]. Based on the data collection methodology of MAP-Neo \cite{zhang2024map}, we collected diverse high-quality data, including: \textbf{1) Natural language data}: FLAN v2 multilingual instruction-tuning dataset \cite{longpre2023flan}, UltraTextbooks high-quality textbooks \cite{UltraTextbooks}, and multi-turn dialogue data cleaned from open networks, enhancing the model's instruction-following and conversational abilities. \textbf{2) Code data}: The Stack dataset (v1) \cite{kocetkov2022stack}, CodeGPT \cite{codegpt}, and LeetCode datasets, covering various programming languages such as Python, C++, and Java. \textbf{3) Mathematics data}: AutoMathText dataset \cite{zhang2024autonomous} and open-web-math dataset \cite{paster2023openwebmath}, improving the model's mathematical problem-solving and logical reasoning abilities.

\subsection{Supervised Fine-Tuning}

In the Supervised Fine-Tuning stage, we refine the large language model to produce high-quality comments for Bash scripts. For this purpose, we employ a token-level negative log-likelihood loss optimized over the dataset:

\begin{equation}
\mathcal{L}_{\mathrm{SFT}} = -\frac{1}{|\mathcal{D}_{\mathrm{exp}}|} \sum_{(x, y) \in \mathcal{D}_{\mathrm{exp}}} \sum_{t=1}^{|y|} \log P_{\theta}(y_t \mid x, y_{<t})
\end{equation}

Here, $x$ is the input Bash script, $y$ the target comment sequence, and $\theta$ denotes all trainable parameters of the LLM (including attention, feed-forward layers, embeddings, etc.).

This objective enables the model to learn to generate comments that accurately and fluently explain the functional and contextual aspects of Bash scripts. The prior domain-specific knowledge acquired during continual pre-training helps the model better understand Bash syntax and semantics, thus supporting the generation of detailed and correct comments.

\subsection{Syntax-Aware Preference Optimization (SAPO)}
In the final stage, to bridge the gap between a generally capable model and a domain expert, we introduce our novel framework: \textbf{Syntax-Aware Preference Optimization (SAPO)}. SAPO reframes the preference optimization problem for code by moving beyond generic quality signals. Instead, it leverages the highly structured and targeted \textbf{``Minimal Syntactic Pairs''} (as generated in Section~\ref{sec:sapo_data}) to inject deep, granular domain knowledge directly into the model.

While SAPO utilizes the same underlying mathematical formulation as Direct Preference Optimization (DPO)~\cite{rafailov2023direct}, its core innovation lies not in the algorithm itself, but in the \textbf{semantic richness of the preference data \textit{D}}. The SAPO objective is to optimize a policy $\pi_{\theta}$ that best satisfies a set of preferences $(x, y_p, y_n)$.

The core of SAPO is based on the insight that the optimal policy $\pi^*$ for a reward function $r^*$ under a KL-constrained optimization objective can be expressed as:
\begin{equation}
\pi^*(y|x) = \frac{1}{Z(x)} \pi_{\text{ref}}(y|x) \exp \left( \frac{1}{\beta} r^*(x, y) \right)
\end{equation}
where $Z(x)$ is a normalization factor, $\pi_{\text{ref}}$ is a reference policy, and $\beta$ is a temperature parameter. In our Bash comment generation scenario:
\begin{itemize}
    \item $x$: represents the input Bash script.
    \item $y$: represents the LLM-generated comment.
    \item $\pi_{\text{ref}}$: the reference policy, typically the model from the SFT stage.
    \item $\pi^*$: the final optimized model.
\end{itemize}

By rearranging this equation, we can express the reward function in terms of the optimal policy:
\begin{equation}
r^*(x, y) = \beta \log \frac{\pi^*(y|x)}{\pi_{\text{ref}}(y|x)} + \beta \log Z(x)
\end{equation}

This allows us to reformulate the Bradley-Terry preference model in terms of policies rather than rewards:
\begin{equation}
p^*(y_p \succ y_n | x) = \sigma \left( \beta \log \frac{\pi^*(y_p|x)}{\pi_{\text{ref}}(y_p|x)} - \beta \log \frac{\pi^*(y_n|x)}{\pi_{\text{ref}}(y_n|x)} \right)
\end{equation}
where $\sigma$ is the logistic function, and $y_p$ and $y_n$ represent the preferred and non-preferred comment outputs.

To simplify, we introduce an intermediate variable $\Delta(x, y_p, y_n)$ representing the log ratio difference:
\begin{equation}
\Delta(x, y_p, y_n) = \beta \log \frac{\pi_{\theta}(y_p|x)}{\pi_{\text{ref}}(y_p|x)} - \beta \log \frac{\pi_{\theta}(y_n|x)}{\pi_{\text{ref}}(y_n|x)}
\end{equation}

Using this, the SAPO loss function can be expressed as:
\begin{equation}
\mathcal{L}_{\text{SAPO}}(\pi_{\theta}; \pi_{\text{ref}}) = -\mathbb{E}_{(x,y_p,y_n) \sim D} \log \sigma(\Delta(x, y_p, y_n))
\end{equation}
where $(x, y_p, y_n)$ are triples from our curated dataset $D$.

In the context of SAPO, this optimization is significantly enhanced. By training on minimal syntactic pairs, $\mathcal{L}_{\text{SAPO}}$ directly maximizes the model's ability to distinguish between correct and incorrect interpretations of specific syntactic elements (e.g., command-line flags or quoting conventions), rather than optimizing for vague ``betterness.'' This transforms DPO into a precise tool for \textbf{targeted knowledge injection}.

\subsection{Prompt Design for LLM Inference}

To ensure fair reproducibility, we standardized prompt designs across all baseline evaluations (GPT-4.1, Claude-3.7-Sonnet, DeepSeek-R1 series, Qwen2.5/3 series, LLaMA-3.1-8B-Instruct) using greedy decoding (temperature=0). \textbf{For Single-line Command Prompting}, following prior work~\cite{shen2024bash, zhang2025bash}, we employ 3-shot prompting by selecting three diverse examples covering file operations, filtering, and conditionals; crucially, these examples are strictly drawn from the training set only to prevent data leakage and ensure no overlap with the test set. Conversely, \textbf{for Multi-line Script Prompting}, we adopt zero-shot prompting to maximize context window utilization, directly instructing models to analyze the script's purpose and workflow.






\subsection{Human Evaluation and LLM Evaluation of Bash Comments}
\label{subsubsec:evaluation_criteria}

\textbf{Human Evaluation.} For human evaluation, following \cite{zhang2025bash, dong2023bash, shen2024bash}, we recruited 24 professional evaluators with over 5 years of Bash programming experience and good English reading skills, divided into 12 groups of 2. We randomly sampled 300 code snippets (200 single-line commands and 100 multi-line scripts) from the test set, assigning 25 unique samples per group. Both group members independently annotated their samples, with a 20\% cross-group overlap (5 samples) ensuring some samples were evaluated by four evaluators. Each evaluator annotated 30 samples in total, with a maximum of 15 samples per 8-hour workday to avoid fatigue. Evaluators were blinded to comment sources, and presentation order and format were randomized and standardized. Inter-evaluator agreement was measured using Fleiss' Kappa over all overlapping samples (rather than averaging group values), yielding 0.71, confirming evaluation reliability. Discrepancies were resolved through group discussion. Each comment was rated on a 1--4 Likert scale across three metrics: Correctness (accurate reflection of code functionality and intent), Completeness (coverage of all key elements without omission), and Naturalness (fluency and readability of language).


\textbf{LLM Evaluation.} To complement human evaluation, we employed \textbf{DeepSeek-R1} as the LLM judge, validated on 100 diverse samples where it achieved \textbf{93\% agreement} with human experts, outperforming GPT-4.1 (89.0\%), Claude-3.7-Sonnet (87\%), LLaMA-3.1-70B (84\%), and Qwen2.5-72B (83.0\%). This performance is attributed to its reinforcement learning-based training~\cite{guo2025deepseek}, which strengthens reasoning in technical documentation tasks. Following LLM-as-a-Judge best practices~\cite{wang2025can, li2025generation, yu2025smart, gu2024survey}, we designed a rigorous evaluation prompt covering \textit{correctness, completeness, and naturalness}, incorporating precise task definitions, granular 1--4 scale scoring criteria, three few-shot anchor examples, and structured JSON output constraints.

\section{Experiments}
\subsection{Research Questions}
To evaluate our Bash-Commenter, we conduct experiments to answer the research questions:

$\bullet$ $\textbf{RQ1}$: How does Bash-Commenter perform in the Bash comment generation task compared to existing methods?

$\bullet$ $\textbf{RQ2}$: How effective are the individual components of Bash-Commenter (CPT, SFT, and SAPO)?

$\bullet$ $\textbf{RQ3}$: How effective are the comments generated by Bash-Commenter in terms of Correctness, Completeness, and Naturalness, as measured by human evaluation and LLM evaluation?


$\bullet$ $\textbf{RQ4} $: What are the characteristic error patterns of Bash-Commenter on scripts of varying complexity?







\subsection{Dataset}

Our dataset comprises continual pre-training (CPT), supervised fine-tuning (SFT), syntax-aware preference optimization (SAPO), and evaluation sets. \textbf{CPT} uses 676,524 Bash scripts from GitHub and Linux documentation (554.40M tokens) plus 300,000 instances from general code, mathematics, and bilingual text (255.39M tokens), totaling 976,524 instances (809.79M tokens). \textbf{SFT} combines 8,469 single-line commands from BASHEXPLAINER \cite{yu2022bashexplainer} and 8,154 multi-line scripts from Dong et al. \cite{dong2023bash}, totaling 16,623 high-quality pairs (5.43M tokens). \textbf{SAPO} contains 2,034 preferred/rejected pairs: 748 from BASHEXPLAINER and 1,286 newly annotated from Dong et al., totaling 1.13M tokens with no SFT overlap. \textbf{Evaluation} uses 2,110 Bash instances: 1,064 single-line commands from BASHEXPLAINER and 1,046 multi-line scripts from Dong et al., with no training overlap. Human evaluation and LLM Evaluation uses 200 single-line commands and 100 multi-line scripts.

\subsection{Baselines and Metrics}

\textbf{Baselines.} Our evaluation covers five categories of baselines: \textbf{Information retrieval-based methods} retrieve comments from similar code snippets based on semantic or lexical similarity, such as LSI~\cite{haiduc2010supporting}, VSM~\cite{haiduc2010use}, BM25~\cite{zhang2020retrieval}, and NNGen~\cite{liu2018neural}. \textbf{Neural network-based methods} generate comments using sequence-to-sequence models, treating the problem as neural machine translation. Examples include CopyNet~\cite{gu2016incorporating}, Transformer~\cite{vaswani2017attention}, CODE-NN~\cite{iyer2016summarizing}, and HBCom~\cite{zhang2025bash}. \textbf{Pre-trained model-based methods} leverage models like CodeBERT~\cite{feng2020codebert}, UniXcoder~\cite{guo2022unixcoder}, CoTexT~\cite{phan2021cotext}, PLBART~\cite{ahmad2021unified}, CodeT5~\cite{wang2021codet5}, BASHEXPLAINER~\cite{yu2022bashexplainer}, and Bash2Com~\cite{shen2024bash}, which are pre-trained on large code corpora and fine-tuned on Bash data. \textbf{Hybrid methods} combine retrieval and neural techniques, as seen in Hybrid-DeepCom~\cite{hu2020deep} and Rencos~\cite{zhang2020retrieval}, aiming to integrate retrieval results into generation. \textbf{LLM-based methods} employ general-purpose LLMs such as GPT-4.1 (gpt-4.1-2025-04-14)~\cite{openai_gpt4o_2024}, Claude-3.7-Sonnet (claude-3-7-sonnet-20250219)~\cite{anthropic2024claude}, LLaMA-3.1-8B-Instruct~\cite{grattafiori2024llama}, Qwen2.5 \cite{yang2024qwen2} and Qwen3 series~\cite{yang2025qwen3technicalreport}, and DeepSeek-R1-Distill-Qwen models~\cite{liu2024deepseek}.

\textbf{Metrics.} We evaluated our model from two dimensions: automatic metrics and human evaluation. For automatic evaluation, we used BLEU-1/2/3/4 (n-gram precision), METEOR (precision/recall with synonyms and stemming), and ROUGE-L (longest common subsequence). Additionally, we employed three key metrics: Correctness, Completeness, and Naturalness (Section~\ref{subsubsec:evaluation_criteria}).

\begin{table*}[ht]
\centering
\caption{Comparison results between our proposed method Bash-Commenter and baselines on Single-line Bash Commands. DeepSeek-R1-7B/14B/32B refer to the distilled versions. LLaMA-3.1-8B and Qwen2.5-7B refer to LLaMA-3.1-8B-Instruct and Qwen2.5-7B-Instruct respectively. MET denotes METEOR metric.}
\label{single_line_comparison}
\renewcommand{\arraystretch}{1.1} 
\setlength{\tabcolsep}{1pt} 
\begin{tabular}{llcccccc}
\toprule
\textbf{Method Type}       & \textbf{Method Name}  & \textbf{BLEU-1} & \textbf{BLEU-2} & \textbf{BLEU-3} & \textbf{BLEU-4} & \textbf{MET} & \textbf{ROUGE-L} \\ 
\midrule
\multirow{4}{*}{Retrieval-based} 
                           & LSI                  & 30.18                & 18.07                & 12.48                & 9.40                 & 18.30                & 28.82                \\ 
                           & VSM                  & 36.16                & 24.47                & 18.62                & 15.25                & 22.04                & 34.58                \\ 
                           & BM25                 & 42.08                & 30.41                & 23.58                & 19.24                & 26.35                & 38.49                \\ 
                           & NNGen                & 50.62                & 38.75                & 32.11                & 27.85                & 27.69                & 45.88                \\ 
\midrule
\multirow{3}{*}{Deep Learning} 
                           & CopyNet              & 38.11                & 27.06                & 20.67                & 16.43                & 22.06                & 40.18                \\ 
                           & Transformer          & 46.39                & 33.37                & 25.42                & 19.97                & 25.22                & 44.01                \\ 
                           & CODE-NN              & 49.60                & 37.18                & 29.53                & 24.17                & 26.85                & 47.21                \\ 
                           & HBCom              & 56.45                & 44.30                & 37.52                & 32.76                & 30.18                & 53.44                \\ 
\midrule
\multirow{7}{*}{Pre-trained Methods} 
                           & CodeBERT             & 48.65                & 37.02                & 29.84                & 24.83                & 27.16                & 47.36                \\ 
                           & UniXcoder            & 49.99                & 38.52                & 31.80                & 27.25                & 29.03                & 48.24                \\ 
                           & CoTexT               & 49.17                & 37.29                & 30.36                & 25.75                & 28.56                & 48.00                \\ 
                           & PLBART               & 50.79                & 39.10                & 32.21                & 27.55                & 28.82                & 47.91                \\ 
                           & CodeT5               & 51.75                & 40.04                & 33.25                & 28.70                & 29.49                & 48.36                \\ 
                           & BASHEXPLAINER        & 51.74                & 40.41                & 33.73                & 29.13                & 28.78                & 48.81                \\ 
                           & Bash2Com             & 54.74                & 43.79                & 37.19                & 32.57                & 30.26                & 51.80                \\ 
\midrule
\multirow{2}{*}{Hybrid Method} 
                           & Hybrid-DeepCom       & 47.78                & 35.45                & 27.91                & 22.75                & 26.27                & 45.36                \\ 
                           & Rencos               & 46.27                & 35.11                & 28.66                & 24.39                & 25.82                & 45.06                \\ 
\midrule
\multirow{9}{*}{General LLMs} 
                           & GPT-4.1               & 20.32                    & 8.16                   & 4.31                    & 2.67                    & 27.67                    & 19.64                    \\ 
                           & Claude-3.7-Sonnet    & 13.62                    & 5.04                    & 2.72                   & 1.74                    & 25.03                    & 15.07                    \\ 
                           & DeepSeek-R1-7B    & 23.15                    & 10.53                    & 5.91                    & 4.10                    & 23.24                   & 24.27    \\ 
                           & DeepSeek-R1-14B    & 25.03                    & 12.59                    & 7.20                    & 4.85                    & 26.98                   & 27.28    \\ 
                           & DeepSeek-R1-32B    & 26.35                    & 13.23                    & 7.48                    & 5.03                    & 28.53                   & 28.92                    \\ 
                           & LLaMA-3.1-8B & 28.37                    & 13.99                    & 7.83                    & 5.12                    & 28.63                    & 30.33                    \\ 
                           & Qwen2.5-7B  & 25.68                    & 13.10                    & 7.48                    & 4.96                    & 28.31                    & 27.96                    \\ 
                           & Qwen3-8B  & 15.64                    & 6.68                    & 3.59                    & 2.22                    & 22.74                    & 17.19                    \\ 
\midrule
\textbf{Our Method}        & \textbf{Bash-Commenter} & 56.21       & \textbf{46.63}       & \textbf{39.65}       & \textbf{33.40}       & \textbf{58.26}       & \textbf{57.03}       \\ 
\bottomrule
\end{tabular}
\end{table*}

\begin{table}[htbp]
\centering
\caption{Performance Comparison on Multi-line Bash Scripts.}
\label{multiline_comparison}
\begin{tabular}{lcccccc}
\toprule
\textbf{Model} & \textbf{B-1} & \textbf{B-2} & \textbf{B-3} & \textbf{B-4} & \textbf{MET} & \textbf{R-L} \\
\midrule
GPT-4.1 & 42.84 & 26.65 & 17.70 & 11.89 & 35.83 & 23.40 \\
Claude-3.7-Sonnet & 44.49 & 26.74 & 17.44 & 11.50 & 34.53 & 22.69 \\
LLaMA-3.1-8B-Instruct & 35.97 & 22.58 & 15.09 & 10.11 & 31.08 & 20.91 \\
DeepSeek-R1-7B & 45.57 & 26.76 & 16.65 & 10.54 & 33.23 & 22.15 \\
DeepSeek-R1-32B & 45.63 & 27.84 & 18.16 & 12.03 & 34.79 & 22.86 \\
Qwen2.5-7B-Instruct & 40.74 & 25.02 & 16.66 & 11.26 & 33.18 & 20.73 \\
Qwen2.5-72B-Instruct & 28.79 & 18.99 & 13.22 & 9.19 & 39.56 & 21.37 \\
Qwen2.5-Coder-32B & 34.45 & 22.89 & 15.96 & 11.17 & 41.93 & 23.55 \\
Qwen3-8B & 30.96 & 18.87 & 12.19 & 7.93 & 36.89 & 17.72 \\
Qwen3-14B & 29.00 & 17.85 & 11.65 & 7.65 & 37.18 & 18.09 \\
\midrule
\textbf{Bash-Commenter} & \textbf{53.11} & \textbf{38.26} & \textbf{28.99} & \textbf{22.15} & \textbf{43.89} & \textbf{32.80} \\
\bottomrule
\end{tabular}
\begin{minipage}{\columnwidth}
\vspace{1mm}
\small Note: B-1/2/3/4=BLEU-1/2/3/4, MET=METEOR, R-L=ROUGE-L. All values are in \%. 
\end{minipage}
\end{table}

\subsection{Implementation Details}
\label{sec:implementation}


All models (CPT, SFT, SAPO) are trained using \texttt{LlamaFactory} and DeepSpeed with fp16, cross-entropy loss, and AdamW optimizer ($\beta=(0.9, 0.99)$, $\epsilon=10^{-8}$), with full parameter tuning. Training is conducted on 8 NVIDIA H800 GPUs (80GB memory each). For CPT: batch size 64, gradient accumulation 16, 3 epochs, learning rate $1\times 10^{-5}$ (cosine decay, no warmup), cutoff length 2048, saving every 500 steps. For SFT: batch size 8, gradient accumulation 8, 3 epochs, same learning rate schedule, cutoff length 2048, saving every 200 steps. For SAPO: batch size 8, gradient accumulation 4, learning rate $1\times 10^{-5}$ (cosine decay, no warmup), cutoff length 1024, 1 epoch. Evaluation uses greedy decoding (temperature $=0$). For baselines of general-purpose LLMs, following previous work \cite{shen2024bash, zhang2025bash}, we use few-shot (3-shot) prompting for single-line commands. For multi-line scripts, we use zero-shot prompting since they are too long for few-shot examples.

\subsection{Experimental Results}


\subsubsection{RQ1. Overall Performance}


We compared Bash-Commenter's performance against baselines.

\paragraph{Quantitative Results}
\textbf{For single-line commands (TABLE \ref{single_line_comparison})}, Bash-Commenter demonstrated strong performance. It achieved a BLEU-4 of 33.40\% (vs. HBCom's 32.76\%) and its ROUGE-L was 57.03\%, surpassing HBCom (53.44\%). Notably, its METEOR score was an exceptionally high 58.26\%, significantly outperforming Bash2Com (30.26\%). This substantial METEOR score is primarily attributed to our three-stage post-training process (CPT, SFT, and SAPO), which greatly enhances the model's ability to capture semantic nuances and generate fluent, contextually relevant comments. Additionally, using LLaMA-3.1-8B as the base model further enhances performance due to its stronger semantic understanding compared to earlier pre-trained models like CodeBERT. \textbf{For multi-line scripts (TABLE \ref{multiline_comparison})}, Bash-Commenter also outperformed other compared LLMs. It achieved a BLEU-4 of 22.15\% (vs. DeepSeek-R1-32B's 12.03\%). Its METEOR was 43.89\% (vs. Qwen2.5-Coder-32B's 41.93\%), and ROUGE-L was 32.80\% (vs. Qwen2.5-Coder-32B's 23.55\%). Earlier CodeBERT-based methods (BASHEXPLAINER, Bash2Com) and deep learning approaches (HBCom) were not used as baselines due to token limits (512 for CodeBERT) and less suitable architectures for complex scripts.

\paragraph{Case Study Analysis}
To complement the quantitative results, we conducted case studies across four representative scenarios.
\textbf{In the first command}, GPT-4.1's comment fails to recognize the specific regex matching pattern and ignores the critical safety parameters (\texttt{-print0} and \texttt{xargs -0}) designed to handle filenames with spaces; BASHEXPLAINER and Bash2Com similarly overlook this mechanism and contain spelling errors (``directori'').
HBCom incorrectly describes the regular expression type as ``posix grep'' instead of the actual ``posix-egrep'', showing a misunderstanding of the command parameters.
In contrast, Bash-Commenter accurately identifies both the regex type and matching pattern, while specifically highlighting the filename space handling safety mechanism.
\textbf{For the second command}, BASHEXPLAINER's comment misses the primary functionality: UUID generation, describing the command as listing filenames.
Bash2Com recognizes the UUID feature but omits sorting and output format details.
HBCom severely misinterprets the command, incorrectly claiming it calculates file sizes and redirects results to a file, neither of which exist.
GPT-4.1 provides a technically accurate comment but focuses excessively on processing mechanisms over practical purpose.
Bash-Commenter accurately captures the core purpose and result organization, specifically noting that results are sorted by UUID value.
\textbf{In the third command}, BASHEXPLAINER and Bash2Com fail to accurately convey the permission setting mechanism, HBCom's ``under and below directory'' is vague and imprecise, and GPT-4.1 overemphasizes the search process with ``recursively find'', making it verbose.
Bash-Commenter concisely describes the permission modification purpose while correctly identifying it operates on regular files in the directory tree.
\textbf{For the fourth command}, BASHEXPLAINER and HBCom both miss the dual filtering mechanism (\texttt{grep -P} and \texttt{grep -v -P}), describing only single filtering, which is a significant functional omission.
Bash2Com captures both filters but omits that decompression outputs to standard output, while GPT-4.1 is complete but verbose.
Bash-Commenter achieves optimal balance by accurately capturing decompression, dual filtering, and line limit while maintaining conciseness.

\vspace{5pt}
\noindent\begin{tikzpicture}
  \node[draw=black, thick, fill=gray!20, rounded corners, inner sep=10pt, text width=0.92\linewidth] {
    \textbf{Answer to RQ1:} Bash-Commenter outperformed existing state-of-the-art methods across main automatic evaluation metrics (BLEU-4, METEOR, and ROUGE-L) for both single-line Bash commands and multi-line Bash scripts. Case studies further demonstrate that these quantitative improvements stem from superior semantic understanding of command-specific behaviors (e.g., regex types, safety mechanisms).
  };
\end{tikzpicture}
\vspace{5pt}

\subsubsection{RQ2. Ablation Study}

Our ablation study (TABLE \ref{ablation_study}) demonstrates the crucial roles of SAPO and CPT. \textbf{For single-line Bash commands}, removing SAPO training significantly degraded performance: BLEU-4 decreased by 10.36 percentage points, ROUGE-L by 10.10, and METEOR by 7.83, indicating SAPO's crucial role in refining comment generation. Removing CPT had greater impact, with BLEU-4 dropping by 14.16 points, ROUGE-L by 11.40, and METEOR by 8.07, highlighting the importance of domain-specific knowledge acquired during this phase. Without both, we observed the most severe degradation: BLEU-4 declined by 17.01 points, ROUGE-L by 16.03, and METEOR by 12.80, demonstrating their complementary nature. \textbf{For multi-line Bash scripts}, performance impacts were less dramatic but significant. Without SAPO training, BLEU-4 decreased by 1.41 points, ROUGE-L by 0.13, and METEOR by 3.89. CPT removal had greater influence, reducing BLEU-4 by 2.96 points, ROUGE-L by 3.16, and METEOR by 4.39. Removing both decreased BLEU-4 by 3.10 points, ROUGE-L by 3.00, and METEOR by 3.47. This reveals a context-dependent pattern: while both components significantly enhance performance for simpler single-line commands, CPT emerges as more critical for complex multi-line scripts, suggesting that domain expertise becomes increasingly paramount as code complexity increases.


\begin{table}[htbp]
\centering
\caption{Ablation Study of Bash-Commenter on Single-line Bash Commands and Multi-line Bash Scripts.}
\label{ablation_study}
\renewcommand{\arraystretch}{1.1}
\setlength{\tabcolsep}{5pt}
\begin{tabular}{llccccc}
\toprule
\textbf{Scenario} & \textbf{Metric} & \textbf{Base} & \textbf{w/o SAPO} & \textbf{w/o CPT} & \textbf{w/o Both} \\
\midrule
\multirow{6}{*}{Single-line} 
 & BLEU-1 & \textbf{56.21} & 46.04 & 42.12 & 37.09 \\
 & BLEU-2 & \textbf{46.63} & 35.60 & 32.75 & 28.40 \\
 & BLEU-3 & \textbf{39.65} & 28.58 & 25.40 & 21.64 \\
 & BLEU-4 & \textbf{33.40} & 23.04 & 19.24 & 16.39 \\
 & METEOR & \textbf{58.26} & 50.43 & 50.19 & 45.46 \\
 & ROUGE-L & \textbf{57.03} & 46.93 & 45.63 & 41.00 \\
\midrule
\multirow{6}{*}{Multi-line}
 & BLEU-1 & \textbf{53.11} & 51.62 & 49.50 & 49.65 \\
 & BLEU-2 & \textbf{38.26} & 36.67 & 34.61 & 34.75 \\
 & BLEU-3 & \textbf{28.99} & 27.43 & 25.62 & 25.61 \\
 & BLEU-4 & \textbf{22.15} & 20.74 & 19.19 & 19.05 \\
 & METEOR & \textbf{43.89} & 40.00 & 39.50 & 40.42 \\
 & ROUGE-L & \textbf{32.80} & 32.67 & 29.64 & 29.80 \\
\bottomrule
\end{tabular}
\end{table}

\vspace{5pt}
\noindent\begin{tikzpicture}
  \node[draw=black, thick, fill=gray!20, rounded corners, inner sep=10pt, text width=0.92\linewidth] {
    \textbf{Answer to RQ2:} Our findings confirm that both CPT and SAPO aid performance with context-dependent importance. CPT offers domain knowledge and has greater impact, especially on multi-line scripts, while SAPO enhances comment quality.
  };
\end{tikzpicture}
\vspace{5pt}

\begin{table}[htbp]
\centering
\caption{Human and LLM Ratings of Correctness, Completeness, and Naturalness.}
\label{evaluation of explanation}
\setlength{\tabcolsep}{3.7pt}
\begin{tabular}{|l|cccc|cccc|cccc|}
\hline
\multirow{2}{*}{\textbf{Model}} & \multicolumn{4}{c|}{\textbf{Correctness}} & \multicolumn{4}{c|}{\textbf{Completeness}} & \multicolumn{4}{c|}{\textbf{Naturalness}} \\
\cline{2-13}
& 1 & 2 & 3 & 4 & 1 & 2 & 3 & 4 & 1 & 2 & 3 & 4 \\
\hline
\multicolumn{13}{|c|}{\textbf{Single-Line Bash Commands (Human Evaluation)}} \\
\hline
BEXPLAIN & 26 & 64 & 73 & 37 & 22 & 65 & 83 & 30 & 14 & 62 & 94 & 30 \\
B2Com & 20 & 57 & 77 & 46 & 18 & 61 & 87 & 34 & 10 & 52 & 99 & 39 \\
HBCom & 20 & 47 & 78 & 55 & 16 & 55 & 93 & 36 & 8 & 48 & 102 & 42 \\
Ours & 7 & 34 & 50 & 109 & 4 & 32 & 58 & 106 & 5 & 17 & 76 & 102 \\
\hline
\multicolumn{13}{|c|}{\textbf{Single-Line Bash Commands (LLM Evaluation)}} \\
\hline
BEXPLAIN & 29 & 66 & 71 & 34 & 25 & 67 & 80 & 28 & 16 & 64 & 92 & 28 \\
B2Com & 23 & 59 & 75 & 43 & 21 & 62 & 85 & 32 & 13 & 52 & 97 & 38 \\
HBCom & 22 & 51 & 76 & 51 & 18 & 58 & 91 & 33 & 11 & 49 & 101 & 39 \\
Ours & 10 & 37 & 52 & 101 & 7 & 33 & 60 & 100 & 7 & 20 & 78 & 95 \\
\hline
\multicolumn{13}{|c|}{\textbf{Multi-Line Bash Scripts (Human Evaluation)}} \\
\hline
GPT-4.1 & 8 & 23 & 41 & 28 & 11 & 20 & 41 & 28 & 4 & 24 & 33 & 39 \\
Claude-3.7 & 9 & 25 & 39 & 27 & 9 & 22 & 43 & 26 & 5 & 23 & 38 & 34 \\
DeepSeek & 13 & 27 & 37 & 23 & 14 & 23 & 40 & 23 & 11 & 18 & 42 & 29 \\
Qwen2.5 & 15 & 27 & 37 & 21 & 13 & 23 & 39 & 25 & 13 & 19 & 40 & 28 \\
Ours & 7 & 20 & 42 & 31 & 11 & 18 & 42 & 29 & 5 & 19 & 42 & 34 \\
\hline
\multicolumn{13}{|c|}{\textbf{Multi-Line Bash Scripts (LLM Evaluation)}} \\
\hline
GPT-4.1 & 10 & 24 & 39 & 27 & 13 & 21 & 40 & 26 & 6 & 25 & 34 & 35 \\
Claude-3.7 & 11 & 26 & 38 & 25 & 11 & 23 & 42 & 24 & 7 & 24 & 37 & 32 \\
DeepSeek & 15 & 28 & 36 & 21 & 16 & 24 & 39 & 21 & 13 & 20 & 40 & 27 \\
Qwen2.5 & 17 & 28 & 36 & 19 & 15 & 24 & 38 & 23 & 15 & 21 & 38 & 26 \\
Ours & 9 & 21 & 41 & 29 & 13 & 19 & 41 & 27 & 7 & 20 & 41 & 32 \\
\hline
\end{tabular}
\begin{minipage}{\columnwidth}
\vspace{1mm}
\small Note: BEXPLAIN: BASHEXPLAINER, B2Com: Bash2Com, Claude-3.7: Claude-3.7-Sonnet, DeepSeek: DeepSeek-R1-Distill-Qwen-32B, Qwen2.5: Qwen2.5-Coder-32B.
\end{minipage}
\end{table}

\subsubsection{RQ3. Human and LLM Evaluation}

TABLE \ref{evaluation of explanation} presents the human and LLM evaluation results. For single-line Bash commands, Bash-Commenter significantly outperformed all baselines in both human and LLM evaluations. In human evaluation, positive ratings (scores 3 and 4) for Correctness, Completeness, and Naturalness reached 79.5\%, 82.0\%, and 89.0\%, respectively. In comparison, HBCom achieved 66.5\%, 64.5\%, and 72.0\%; Bash2Com scored 61.5\%, 60.5\%, and 69.0\%; and BASHEXPLAINER had 55.0\%, 56.5\%, and 62.0\% on the three metrics. LLM evaluation showed consistent trends with positive ratings of 76.5\%, 80.0\%, and 86.5\% for our method, compared to HBCom's 63.5\%, 62.0\%, and 70.0\%; Bash2Com's 59.0\%, 58.5\%, and 67.5\%; and BASHEXPLAINER's 52.5\%, 54.0\%, and 60.0\%. For multi-line Bash scripts, Bash-Commenter also led in human evaluation with Correctness (73.0\%), Completeness (71.0\%), and Naturalness (76.0\%), surpassing GPT-4.1 and Claude-3.7-Sonnet, both of which scored no higher than 69.0\% in Correctness or Completeness, and 72.0\% in Naturalness. LLM evaluation corroborated these findings with our method achieving 70.0\% in Correctness, 68.0\% in Completeness, and 73.0\% in Naturalness, while GPT-4.1 and Claude-3.7-Sonnet remained below 66.0\%, 66.0\%, and 70.0\% respectively.

\vspace{5pt}
\noindent\begin{tikzpicture}
  \node[draw=black, thick, fill=gray!20, rounded corners, inner sep=10pt, text width=0.92\linewidth] {
    \textbf{Answer to RQ3:} Bash-Commenter has demonstrated its capability to generate comments for Bash commands that are correct, complete, and natural validated by both human and LLM evaluations, outperforming all baselines.
  };
\end{tikzpicture}
\vspace{5pt}

\subsubsection{RQ4. Error Analysis}
\label{sec:error_analysis}


To better understand the limitations of Bash-Commenter, we performed a manual error analysis on 100 incorrect comments, comprising 50 single-line commands and 50 multi-line scripts, that received a `Correctness' score of 1 or 2 in our human evaluation; two experts independently categorized the primary error in each comment, resolving disagreements through discussion, into six specific categories: \textbf{Parameter/Option Omission (PO)}, where the model correctly identifies the command but omits key parameter functions; \textbf{Functionality Misinterpretation (FM)}, which misunderstands the core purpose of a command or pipeline; \textbf{Incompleteness (IC)}, denoting comments that are correct at a high level but lack crucial logic details; \textbf{Factual Hallucination (FH)}, where the model invents functionality not present in the script; \textbf{Over-generalization (OG)}, which is generic and fails to capture specific context; and \textbf{Language/Fluency Issues (LFI)}, covering comments containing grammatical errors or awkward phrasing.

\begin{table}[ht]
\centering
\caption{Distribution of Error Types for Bash-Commenter (\%)}
\label{tab:error_distribution_en}
\begin{tabular}{lcc}
\toprule
\textbf{Error Category} & \textbf{Single-line Scripts (\%)} & \textbf{Multi-line Scripts (\%)} \\
\midrule
Parameter/Option Omission (PO)      & 38 & 18 \\
Functionality Misinterpretation (FM)  & 24 & 28 \\
Incompleteness (IC)                 & 16 & 34 \\
Factual Hallucination (FH)          & 8  & 6  \\
Over-generalization (OG)            & 10 & 12 \\
Language/Flsues (LFI)       & 4  & 2  \\
\midrule
\textbf{Total}                      & \textbf{100} & \textbf{100} \\
\bottomrule
\end{tabular}
\end{table}


Our analysis (Table~\ref{tab:error_distribution_en}) identifies \textbf{Parameter/Option Omission (PO)} as the dominant error for \textbf{single-line commands} (38\%), indicating the model grasps the main utility but misses flag-induced nuances. Conversely, \textbf{multi-line scripts} suffer chiefly from \textbf{Incompleteness (IC)} (34\%), reflecting the difficulty of summarizing complex logic steps. To address PO, we propose Group Relative Policy Optimization (GRPO). By evaluating a group of candidates against a reward function, GRPO provides granular signals to master subtle flag impacts. For IC, Chain-of-Thought (CoT) fine-tuning offers a solution by enforcing a step-by-step logical breakdown, ensuring all crucial script components are captured in the final summary.

\vspace{5pt}
\noindent\begin{tikzpicture}
  \node[draw=black, thick, fill=gray!20, rounded corners, inner sep=10pt, text width=0.92\linewidth] {
    \textbf{Answer to RQ4:} Bash-Commenter's main errors are parameter omission (PO) for single-line commands and incompleteness (IC) for multi-line scripts. Future work can address these using GRPO for finer optimization and CoT fine-tuning.
  };
\end{tikzpicture}
\vspace{5pt}

\section{Related Work}

\subsection{Bash Code Generation from Natural Language}


Mapping natural language to Bash commands is challenging. Lin et al.\cite{lin2017program} first addressed this with a corpus of over 9{,}000 Bash code-comment pairs covering 100+ utilities, evaluating Seq2Seq\cite{sutskever2014sequence}, CopyNet~\cite{gu2016incorporating}, and Tellina~\cite{lin2017program}. DocCGen~\cite{pimparkhede2024doccgen} improved NL-to-DSL generation through a two-stage process: retrieving relevant library documentation, then generating code from syntax rules and templates, reducing syntax and semantic errors in complex DSLs like Bash and Ansible YAML. Bridge-Coder~\cite{zhang2024bridge} addressed low-resource language code generation via ``Code-Bridge,'' leveraging code and comments from high-resource languages to guide generation in low-resource languages.



\subsection{Bash Code Comment Generation}

For Bash code comment generation, Yu et al.\cite{yu2022bashexplainer} proposed BASHEXPLAINER via CodeBERT\cite{feng2020codebert}-based encoding and information retrieval, Shen et al.\cite{shen2024bash} introduced Bash2Com with adversarial training and a semantic-aware module, and Zhang et al.\cite{zhang2025bash} proposed HBCom using a Heterogeneous Information Graph for syntactic-semantic integration. Despite this progress, these methods share three limitations: CodeBERT's 512-token ceiling prevents handling long multi-line scripts; datasets cover only hundreds of distinct commands (e.g., 466 in BASHEXPLAINER~\cite{yu2022bashexplainer}), limiting practical diversity; and the absence of parameter-level semantic modeling causes errors such as misinterpreting \texttt{uniq -d} or missing \texttt{awk -F} and \texttt{uniq -i} semantics (Fig.\ref{tab:motivation}). Our work addresses these gaps by constructing a dataset of \textbf{17,122 distinct commands} (36.7×36.7× larger) with 50.6\% multi-line scripts (Table\ref{tab:dataset-comparison}), adopting \textbf{LLaMA-3.1-8B}~\cite{grattafiori2024llama} with a 2048-token context and CPT on 676,524 Bash scripts, and proposing \textbf{SAPO}, the first preference learning framework using AST-based minimal pairs for fine-grained command semantics.

\subsection{Automatic Code Comment Generation}


Code comment generation has evolved from rule-based templates~\cite{haiduc2010use} to neural models like CodeNN~\cite{iyer2016summarizing}. Recent LLM-based approaches have achieved significant breakthroughs: Geng et al.\cite{geng2024large} introduced intent-guided generation; Lu et al.~\cite{lu2025deepcrceval} established the DeepCRCEval framework for reliable evaluation; while others improved performance through parameter-efficient fine-tuning~\cite{lu2023llama}. Additionally, to better capture the structural semantics of source code, researchers have extensively explored graph neural networks (GNNs)~\cite{yang2023dccgraph,zhang2026binary, zhang2025topology, yu2023money} to represent code structures~\cite{kuang2022code}. However, these works primarily focus on general-purpose languages (e.g., Java, Python) rather than domain-specific languages like Bash.

\section{Threats to Validity}



\textbf{Internal Validity}: Internal validity threats include occasional output inconsistencies such as repetitive phrases or unintended inclusion of the input script in generated comments; these are mitigated by rule-based post-processing in the inference pipeline. Inference runs efficiently on a single 24GB VRAM GPU. The risk of systematic bias from LLM-generated reference text is addressed through 24-expert verification (Fleiss' Kappa = 0.78) and 500-sample validation. Future work will explore refined training objectives and architectural adjustments to further improve output quality.

\textbf{External Validity}: SAPO generates preference pairs via atomic mutations applied to a script's AST, but the semantic correctness of these mutations is implicitly learned from training data patterns. If the model lacks exposure to a command's valid argument structure, SAPO cannot construct meaningful correct-vs-incorrect pairs for it. Thus, while SAPO is fully automated, its effectiveness in new or specialized domains remains contingent on sufficient foundational data coverage for that domain.

\section{Conclusion}


We propose Bash-Commenter, an LLM-based approach for Bash code comment generation that employs a three-stage post-training pipeline: continual pre-training (CPT), supervised fine-tuning (SFT), and syntax-aware preference optimization (SAPO), supported by a comprehensive dataset constructed for each stage. Bash-Commenter significantly outperforms state-of-the-art methods on BLEU-4, METEOR, and ROUGE-L, generating comments that are correct, complete, and natural.


\section{Data Availability}

All the experimental data and source code is online available at \url{https://zenodo.org/records/18743947}

\section*{Acknowledgements}
This work was supported by the National Key Research and Development Program of China (No.2023YFB3307203). 

\bibliographystyle{ACM-Reference-Format}
\bibliography{References}
\end{document}